\documentclass[fleqn,twoside]{article}
\usepackage{espcrc2}
\usepackage{graphicx}

\newcommand{\AmS}{{\protect\the\textfont2
  A\kern-.1667em\lower.5ex\hbox{M}\kern-.125emS}}

\newcommand{\be}{\begin{equation}}
\newcommand{\ee}{\end{equation}}
\newcommand{\bea}{\begin{eqnarray}}
\newcommand{\eea}{\end{eqnarray}}
\newcommand{\bi}{\begin{itemize}}
\newcommand{\ei}{\end{itemize}}
\newcommand{\MSbar}{\overline{\rm{MS}}}

\usepackage{amssymb}

\normalsize
\title{
\vspace*{-35pt}
{\normalsize \hfill {\sf KEK-CP-116}} \\
\vspace*{20pt}
Light hadron spectrum and quark masses
}
\author{
T.~Kaneko\address{High Energy Accelerator Research Organization
                     (KEK), Tsukuba, Ibaraki 305-0801, Japan}
} 

\begin{document}

\begin{abstract}
  Recent developments in lattice QCD calculations
  of the light hadron spectrum and quark masses are reviewed.
\end{abstract}

\maketitle
\setcounter{footnote}{0}

\section{Introduction}
\label{sec:intro}


One of the main goals of lattice QCD is to confirm the
validity of QCD as the theory of strong interactions.
Calculation of the hadron spectrum is fundamental in this respect. 
In the quenched approximation, large effort over the years 
has culminated in the work by CP-PACS\cite{CP-PACS.Nf0} 
which revealed a systematic deviation 
of the quenched spectrum from experiment. 
%
%
Since then, the focus has shifted to full QCD including 
dynamical quarks, and a number of large-scale simulations have been 
executed in $N_f\!=\!2$ full QCD\cite{review.full}.  
A crucial issue here is how sea quark effects manifest in 
the spectrum.  Work to explore this question has continued this year.


A realistic QCD simulation has to treat the strange quark dynamically. 
The algorithmic development to treat odd number of flavors has accelerated 
in the last few years\cite{review.algo}.  This has led to several attempts 
toward realistic calculations in $N_f\!=\!3$ QCD very recently.


One of the major uncertainties in lattice calculations
arises from the chiral extrapolation toward realistically 
light quark masses.
One possibility to control this extrapolation is to use 
chiral perturbation theory (ChPT)\cite{ChPT} as a guide.
Test of the validity of ChPT is an important step for this purpose, 
and full QCD data to examine this issue are becoming available.


Another recent trend is the improvement of formulations of lattice QCD 
such as the domain-wall fermion\cite{review.GW} 
and fermions defined on anisotropic lattices\cite{aniso}. 
Quenched QCD presents a testing ground of these 
formulations, and some realistic simulations have been made.


Spectrum calculations allow a simultaneous determination of 
the strong coupling constant and quark masses 
which are 
the fundamental parameters of the Standard Model. 
Precise determination of these quantities is an important 
issue in lattice QCD\cite{review.alphas,review.mq}.
New studies in this area have also been reported at this conference.


In this review, we present the status of 
lattice calculations of the light hadron spectrum and  
the fundamental parameters of QCD with focus on the points above.
We concentrate on results obtained with the Wilson-type quark actions, 
as those with the Kogut-Susskind (KS) action are covered  
in a separate talk\cite{KS.Toussaint}.
%
%
In Sec.~\ref{sec:Nf2},
we discuss sea quark effects in $N_f\!=\!2$ full QCD.
Recent developments toward $N_f\!=\!3$ simulations are reviewed 
in Sec.~\ref{sec:Nf3}, 
and verification of ChPT is discussed in Sec.~\ref{sec:ChPT}.
Section \ref{sec:Nf0} is devoted to recent developments
of spectroscopic studies in quenched QCD.
The status of the strong coupling constant  
and quark masses are updated in Secs.~\ref{sec:alpha_s} 
and \ref{sec:mq}. 
A brief conclusion is given in Sec.~\ref{sec:conclusion}.

\begin{table*}[t]
\caption{
  Recent simulations in $N_f\!=\!2$ QCD.
  Values of $c_{\rm SW}$ for the clover quark action\cite{SWaction}
  are shown in brackets, where NP and TP stand for non-perturbative 
  and tadpole-improved values, respectively.
}
\begin{tabular}{llllllll}
\hline 
\hline 
group            & gauge        & quark                  
                 & $a_s$[fm]    & $L_s$[fm]             
                 & $m_{\rm PS,sea}/m_{\rm V,sea}$ & ref.         \\
\hline
SESAM            & plaquette    & Wilson                 
                 & 0.08         & 1.3          
                 & 0.68--0.83   & \cite{SESAM.Nf2}       \\
T$\chi$L         & plaquette    & Wilson                 
                 & 0.08         & 1.9           
                 & 0.57, 0.70   & \cite{TchiL.Nf2}       \\ 
SESAM-T$\chi$L   & plaquette    & Wilson             
                 & 0.09         & 1.5         
                 & 0.68--0.85   & \cite{SESAM.Nf2.beta55}\\ 
\hline
UKQCD            & plaquette    & clover (1.76)     
                 & 0.12         & 1.0--1.9     
                 & 0.67--0.86   & \cite{UKQCD.Nf2.csw176}\\
UKQCD            & plaquette    & clover (NP) 
                 & 0.10         & 1.7
                 & 0.70--0.84   & \cite{UKQCD.Nf2.lat00,UKQCD.Nf2}\\
UKQCD            & plaquette    & clover (NP) 
                 & 0.10         & 1.6
                 & 0.58         & \cite{UKQCD.Nf2.lat00,UKQCD.Nf2}\\
QCDSF            & plaquette    & clover (NP)      
                 & 0.09         & 2.1,1.5      
                 & 0.69, 0.76   & \cite{QCDSF.Nf2.lat00} \\
JLQCD            & plaquette    & clover (NP)      
                 & 0.09         & 1.1--1.8      
                 & 0.60--0.80   & \cite{JLQCD.Nf2.lat00,JLQCD.Nf2}\\
\hline
CP-PACS          & RG-improved\cite{RGgauge}  
                 & clover (TP)  
                 & 0.11--0.22   & 2.5--2.6      
                 & 0.55--0.81   & \cite{CP-PACS.Nf2}     \\
\hline
\hline
Columbia         & plaquette    & KS
                 & 0.09         & 1.5           
                 & 0.57--0.70   & \cite{Coulumbia.Nf2}   \\
MILC             & plaquette    & KS
                 & 0.10--0.32   & 2.4--3.8      
                 & 0.3--0.8     & \cite{MILC.Nf2.stdKS}  \\
Columbia         & plaquette    & KS ($\xi\!\sim\!1.8$--5.0) 
                 & 0.23--0.34   & 3.7--5.4      
                 & 0.35--0.80   & \cite{Columbia.Nf2.aniso} \\
\hline
MILC             & Symanzik\cite{1loopSym}
                 & improved KS\cite{improvedKS}
                 & 0.13         & 2.6          
                 & 0.50         & \cite{MILC.Nf3}        \\
\hline
\hline
\end{tabular}
\label{tab:simulations.Nf2}
\vspace{-2mm}
\end{table*}

%
%
%
%
%
%
%
%
%

\section{Light hadron spectrum in $N_f\!=\!2$ QCD}
\label{sec:Nf2}

\subsection{Recent simulations}


Recent simulations in two-flavor QCD 
are listed in Table~\ref{tab:simulations.Nf2}.
%
%
Most of the simulations with the Wilson-type quark action 
were made with the plaquette gauge action at a single lattice 
spacing $a\sim 0.1$~fm. 
An exception is the CP-PACS study\cite{CP-PACS.Nf2}
which explored the range $a \! \sim \! 0.1$--0.2~fm with the use of 
an improved gauge action.  
This work reported clear sea quark effects in the meson spectrum 
after the continuum extrapolation.
%
%
%
%

%
%
New results were reported by UKQCD\cite{UKQCD.Nf2} 
and JLQCD\cite{JLQCD.Nf2} at this conference.
The two simulations still use the plaquette gauge action and work 
at a single lattice spacing $a\!\sim\!0.1$~fm, 
but the quark action is fully $O(a)$ improved 
with $c_{\rm SW}$ determined non-perturvatively\cite{NP-csw.ALPHA}.

The UKQCD simulation shifts $\beta$ with sea quark mass, 
keeping the Sommer scale $r_0/a$\cite{r0} fixed. 
The JLQCD runs were performed at a fixed $\beta$.
%
%
Another important difference 
is the range of quark mass covered in the two simulations: 
JLQCD explored light sea quark masses down to 
$m_{\rm PS,sea}/m_{\rm V,sea}\!\sim\!0.6$,
whilst the UKQCD's lightest point is around 
$m_{\rm PS,sea}/m_{\rm V,sea}\!\sim\!0.7$.\footnote[2]{
Although UKQCD made another simulation at a smaller sea quark mass
($m_{\rm PS,sea}/m_{\rm V,sea}\!\sim\!0.58$, $L_s\!\sim\!1.6$~fm),
finite size effects seem to be significant there\cite{JLQCD.Nf2}.}
We note that Orth {\it et al.} attempts to reduce quark masses 
to even lighter points, 
but actual simulations have not yet been started\cite{GRAL.Nf2}.

%
%
%

\subsection{Sea quark effects in meson spectrum}  
         
\begin{figure}[t]
\begin{center}
   \leavevmode
   \includegraphics*[width=66mm,clip]{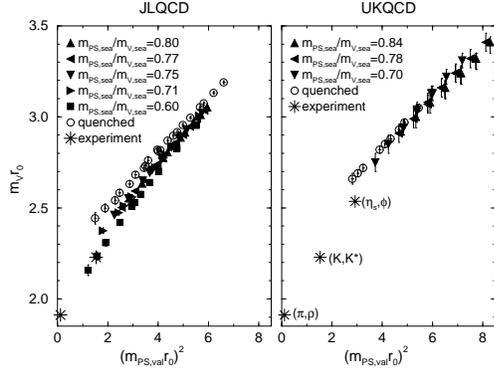}
\end{center}
\vspace*{-15mm}
\caption{
   Vector meson mass as a function of pseudo-scalar meson mass squared,
   normalized in units of $r_0\!=\!0.49$~fm from JLQCD and UKQCD. 
}
\vspace*{-5mm}
\label{fig:Nf2:mVvsmPS2}
\end{figure}

Figure~\ref{fig:Nf2:mVvsmPS2} shows 
how the valence quark mass dependence of the vector meson mass 
varies as a function of sea quark mass 
in the JLQCD and UKQCD simulations.
In both UKQCD and JLQCD data, 
the set of points for the sea quark masses 
down to $m_{\rm PS,sea}/m_{\rm V,sea}\!\sim\!0.7$ almost overlap.  
On the other hand, the JLQCD data for a lighter sea quark 
$m_{\rm PS,sea}/m_{\rm V,sea}\!\sim\!0.6$ clearly show a larger slope, 
and are closer to the experimental points marked by asterisks.  

%
%
%
%

The smallest sea quark mass in most of the previous simulations 
stopped at $m_{\rm PS,sea}/m_{\rm V, sea} \! \sim \! 0.7$, 
which explains why sea quark effects were not clearly seen 
in these simulations.  
%
%
The CP-PACS data did cover the range down to   
$m_{\rm PS,sea}/m_{\rm V, sea} \! \sim \! 0.6$.
The trend of sea quark effects as observed by JLQCD
is present at $a\!\sim\!0.1$~fm,
while it is less clear on coarser lattices 
due to larger error of $r_0$\cite{CP-PACS.mVvsmPS2}. 

%

\begin{figure}[t]
\vspace*{-2mm}
\begin{center}
   \leavevmode
   \includegraphics*[width=65mm,clip]{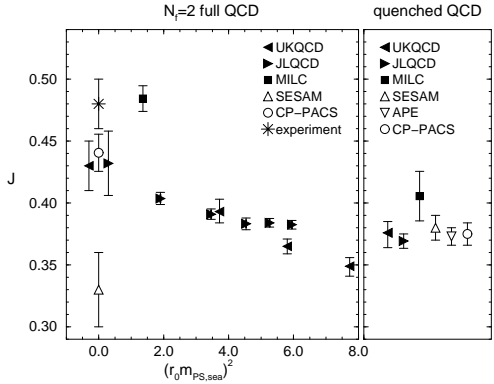}
\end{center}
\vspace*{-15mm}
\caption{
  Recent results of $J$ parameters
  in $N_f\!=\!2$ full QCD and quenched QCD. 
  Results presented this year are plotted 
  in filled symbols.
}
\vspace*{-6mm}
\label{fig:Nf2:J}
\end{figure}

%
%
Figure~\ref{fig:Nf2:J} shows recent results of the $J$ parameter\cite{J}
calculated with differentiation in the valence quark mass 
as a function of the sea quark mass.
%
%
The JLQCD data show 
a trend that the value of $J$, which is consistent with that of quenched 
QCD for heavier sea quark masses, increases as the sea quark mass 
decreases.  The latter feature can be observed also in the UKQCD data.
As a result, 
the values extrapolated to the physical sea quark mass
are closer to experiment than in quenched QCD.

The $J$ parameter in full QCD 
can also be obtained using the experimental definition 
$J \equiv m_{K^*}(m_{K^*}\!-\!m_{\rho})/(m_{K}^2\!-\!m_{\pi}^2)$, 
or from the combined chiral extrapolation as a function of 
sea and valence quark masses.  The results are also plotted at 
$(r_0 m_{\rm PS,sea})^2\!=\!0$ in Fig.~\ref{fig:Nf2:J}.
They lead to the same conclusion 
that $J$ increases with dynamical quarks.
An exception is the SESAM point
presumably due to heavier sea quark masses
$m_{\rm PS,sea}/m_{\rm V,sea}\!\gtrsim\!0.7$
and scaling violation in their simulations.
Recent results with KS fermions also observed
similar sea quark effects\cite{MILC.Nf3}.

\begin{figure}[t]
\begin{center}
   \leavevmode
   \includegraphics*[width=65mm,clip]{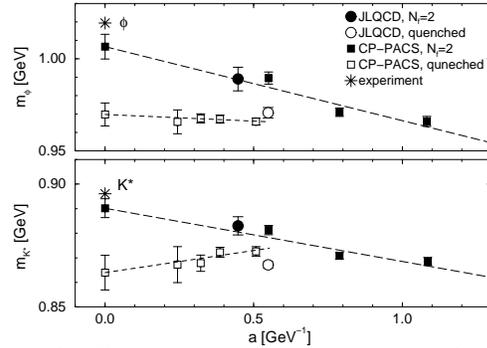}
\end{center}
\vspace*{-16mm}
\caption{
  Strange vector meson masses with $m_K$ as input
  in $N_f\!=\!2$(filled symbols)
  and quenched QCD(open symbols).
}
\vspace*{-6mm}
\label{fig:Nf2:meson}
\end{figure}

The results so far refer to a single lattice spacing 
of $a \! \sim \! 0.1$~fm. In Fig.~\ref{fig:Nf2:meson} 
we reproduce the previous results by CP-PACS\cite{CP-PACS.Nf2}
on the strange vector meson masses 
for a set of lattice spacings, 
and superimpose the new JLQCD points. 
%
%
The JLQCD points are consistent with the CP-PACS points
both for $N_f\!=\!2$ and $N_f\!=\!0$ (quenched), 
showing an increase of hyperfine splitting 
as dynamical quark effects are included.    

Strictly speaking comparison of the two data
should be performed in the continuum limit, albeit 
scaling violation in the JLQCD data may be expected to be small 
by the use of the non-perturbatively $O(a)$-improved action. 
This point should be checked in future studies.

\subsection{Baryons and finite size effects}  

Significant sea quark effects are also expected in the baryon spectrum,
as it sizably deviates from experiment in quenched QCD\cite{CP-PACS.Nf0}.
However, no group has found clear evidence.
A possible reason is finite size effects,
which we expect to be more important for baryons than for mesons. 

\begin{figure}[t]
\begin{center}
   \leavevmode
   \includegraphics*[width=65mm,clip]{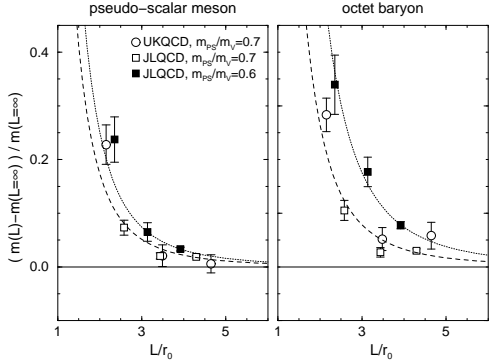}
\end{center}
\vspace*{-15mm}
\caption{
  Mass shift at finite lattice size $L$ as a function of 
  $L/r_0$. 
  The dashed and dotted lines shows fit curve for data 
  at sea quark mass corresponding to $m_{\rm PS}/m_{\rm V}\!=\!0.7$
  and 0.6, respectively.
}
\vspace*{-6mm}
\label{fig:Nf2:FSE}
\end{figure}

Suppression of systematic errors to a few percent level 
is required for a convincing study of sea quark effects.
In order to estimate the lattice size needed 
to achieve this accuracy,
we fit hadron mass data of UKQCD\cite{UKQCD.Nf2.csw176}
and JLQCD\cite{JLQCD.Nf2} on several lattice volumes 
to an ansats $m(L)=m(L\!=\!\infty)+c/L^3$\cite{FSE.1/V}.
Using the fitted value of $m(L\!=\!\infty)$, 
we show in Fig.~\ref{fig:Nf2:FSE} the relative magnitude of 
finite size effects as a function of the spatial extent $L$
(divided by $r_0$)
at sea quark masses corresponding to $m_{\rm PS}/m_{\rm V}=0.7$ 
and 0.6.

We observe that a lattice size of $L \sim 2$~fm suffices 
for mesons to keep finite size effects under a few percent level for 
sea quark masses down to $m_{\rm PS}/m_{\rm V}\sim 0.6$. 
On the other hand, 
calculations of the baryon masses need 
$L \sim 3$~fm at the same sea quark masses 
and the required size increases rapidly
as the sea quark mass decreases.
Further 
studies on such large volumes 
will be necessary to identify sea quark effects on the baryon spectrum.

%

\subsection{Glueball masses}

An unambiguous experimental identification of the glueball 
is still missing, 
and lattice QCD can provide helpful theoretical estimates.
Calculations in quenched QCD have been performed for a long time 
and recent results for $J^{PC}\!=\!0^{++}$ and $2^{++}$ 
states are in good agreement with each other\cite{review.glue.lat99}.
The focus has now shifted to calculations in full QCD
and mixings with quarkonia.
%
%
%

UKQCD\cite{UKQCD.Nf2.glue} and SESAM-T$\chi$L\cite{SESAM.Nf2.glue}
reported new results this year.
Both groups find a decrease of $m_{0^{++}}$ by about 20\%
with dynamical quarks.
The observed sea quark mass dependence, however, is quite different:
SESAM-T$\chi$L finds a significant slope, 
while UKQCD data are almost flat.
It is, thus, too early to draw a definite conclusion
on the sea quark effect.  Scaling violation should be 
also studied carefully, as it is known to be large in 
quenched QCD. 
A similar indication was already 
seen by SESAM-T$\chi$L\cite{SESAM.Nf2.glue} in full QCD.

%

\section{QCD with dynamical up, down  and strange quarks} 
\label{sec:Nf3}

Realistic simulations of QCD require up, down and strange 
quarks to be treated dynamically.  While such studies already exist 
for the KS quark action using an approximate R algorithm\cite{KS.Toussaint,MILC.Nf3}, 
exact algorithms for odd number of Wilson-type quarks
were developed only recently\cite{review.algo}. 
%
%

\begin{figure}[t]
\begin{center}
   \leavevmode
   \includegraphics*[width=65mm,clip]{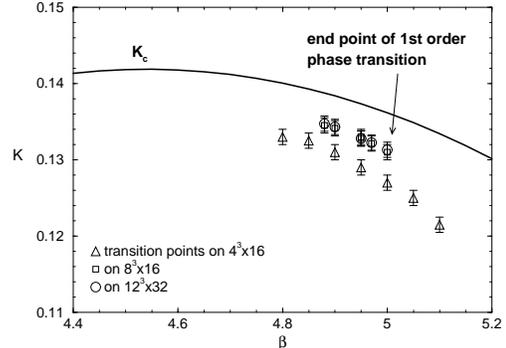}
\end{center}
\vspace*{-15mm}
\caption{
  Phase diagram in $N_f\!=\!3$ full QCD with 
  the plaquette gauge and $O(a)$-improved Wilson quark.
}
\vspace*{-6mm}
\label{fig:Nf3:PD}
\end{figure}

This year JLQCD initiated a study of $N_f=3$ full QCD 
using an exact algorithm 
for the $O(a)$-improved Wilson fermion\cite{JLQCD.Ishikawa}. 
As a first step, they studied the phase structure 
on the $(\beta, \kappa)$ plane 
and found an unexpected first-order phase transition
with the standard plaquette gauge 
and the $O(a)$-improved Wilson quark action,
as shown in Fig.~\ref{fig:Nf3:PD}\cite{JLQCD.Okawa}.
The transition points do not move 
between $8^3\times16$ and $12^3\times32$ lattices, 
and they disappear at $\beta\!\geq\!5.0$.
Hence this is a bulk transition at zero temperature, 
and is a lattice artifact. 
%

Their finding poses a serious practical problem, 
because the lattice spacing estimated from $r_0$ 
equals $a^{-1} \! \sim \! 2.6$~GeV
at the end-point of the first-order transition at $\beta\!\sim\!5.0$.  
Large-scale simulations at such fine lattice spacings are
not practical with the computer power currently available.

They found, however, that this unphysical phase transition 
disappears if improved gauge actions are employed. 
For realistic simulations, 
it is a natural choice to use improved gauge actions 
at moderately large values of lattice spacing such as 
$a \! \sim \! 0.1$--0.2~fm.

%
%

\section{Test of chiral perturbation theory}
\label{sec:ChPT}

%

\begin{figure}[t]
\vspace*{-2mm}
\begin{center}
   \leavevmode
   \includegraphics*[width=66mm,clip]{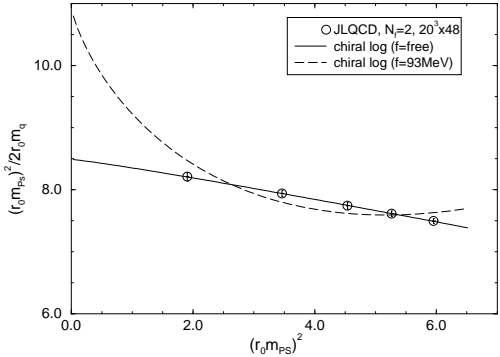}
\end{center}
\vspace*{-15mm}
\caption{
Test of chiral perturbation theory.
}
\vspace*{-6mm}
\label{fig:ChPT:JLQCD}
\end{figure}

Chiral perturbation theory (ChPT)\cite{ChPT} predicts 
the presence of logarithm singularities 
in the quark mass dependence of physical quantities,
{\it e.g.}, for the pseudo-scalar meson mass $m_{\rm PS}$, 
\bea
   \frac{m_{\rm PS}^2}{2B_0m_q}
      &=&
      1 + \frac{1}{N_f} y \ln \left[ y \right] 
      + O(y)
%
\label{eq:ChPT:mPS2}
\eea
where $y \equiv 2B_0m_q/(4\pi f)^2$. 
%
A test of this relation is shown in Fig.~\ref{fig:ChPT:JLQCD},
where lattice data for two-flavor QCD from JLQCD\cite{JLQCD.Nf2}
are plotted together with the fit curves of Eq.~(\ref{eq:ChPT:mPS2}) 
assuming $f$ to be a free parameter or fixed to $f=93$~MeV.
The lattice results are not consistent with the curvature
characterized by the chiral logarithm.

Recently UKQCD\cite{ChPT.Nf2.UKQCD}
attempted to extract the low energy constants
of chiral lagrangian from a fit of lattice data 
to the partially quenched ChPT formula\cite{PQChPT}
using ratios proposed by ALPHA\cite{ChPT.Nf0.ALPHA}.
The curvature expected from the chiral logarithm is also not clear
in their data.

A possible reason behind these disagreements 
is that sea quark masses in  current simulations are too large.  
In fact, 
the coefficient of the chiral logarithm is modified
when the effect of flavor singlet meson is included into 
ChPT\cite{pqChPT.with.eta}; 
it can take much smaller values than $1/N_f$, 
if the pesudo-scalar meson is not so light
compared to the flavor singlet meson\cite{JLQCD.Nf2}.
A recent work by Nelson {\it et al.}\cite{ChPT.Nf3.Nelson}
using KS fermions suggests a similar conclusion.
The curvature is visible in their data at small sea quark masses, 
which are possible with KS fermions.

Another interesting subject of ChPT is 
determination of the low energy constants $\alpha_i$ 
from lattice data\cite{mu0,ChPT.const}.
In particular, 
$\alpha_8$ is an important quantity 
since it is related to the possibility of $m_u\!=\!0$\cite{mu0}.
UKQCD  attempted to extract $\alpha_i$'s in two-flavor QCD, 
and obtained $\alpha_8\!=\!0.79(^{+5}_{-7})(21)$, 
which disfavors $m_u\!=\!0$.
Nelson {\it et al.} obtained a consistent conclusion 
in three flavor QCD.
These results, however, should be confirmed with lighter sea quark masses,
for which ChPT at the one-loop level becomes more reliable.


\section{Developments in quenched QCD}
\label{sec:Nf0}

\subsection{Test of improved formulations}

Quenched QCD is now used to develop and test 
improved formulations of lattice QCD.
Domain-wall fermion formalism is an important case,
as it is expected to 
realize exact chiral symmetry at finite lattice spacings.
Anisotropic lattices may provide a way 
to achieve simulations with large lattice cut-offs 
on large spatial volumes.

\begin{figure}[t]
\begin{center}
   \leavevmode
   \includegraphics*[width=66mm,clip]{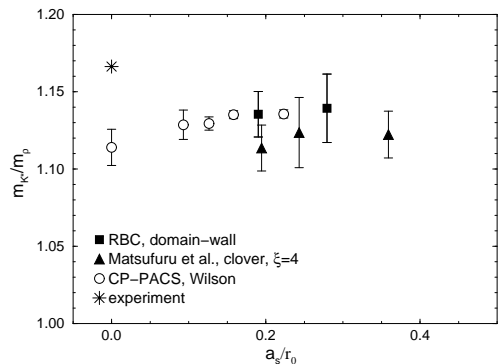}
\end{center}
\vspace*{-15mm}
\caption{
  Continuum extrapolation of $m_{K^*}/m_{\rho}$
  obtained with domain-wall fermion(squares) 
  and on anisotropic lattice(triangles).
}
\vspace*{-6mm}
\label{fig:Nf0:mKs}
\end{figure}

Figure~\ref{fig:Nf0:mKs} shows 
results of $m_{K^*}/m_{\rho}$ mass ratio calculated 
with the domain-wall fermions\cite{DW.RBC},
and on anisotropic lattices\cite{aniso.Matsufuru}.
Clearly these formulations reproduce, and confirm, the small 
hyperfine splitting in the quenched meson spectrum 
obtained by the precison calculation by CP-PACS
using the conventional Wilson quark action\cite{CP-PACS.Nf0}.

Recently, the Wilson quark action with a chirally twisted 
mass term\cite{tmQCD.Aoki}
has drawn attention as an alternative formulation to suppress 
exceptional configurations\cite{tmQCD}.
Some evidences of such suppression 
have been presented 
at this conference\cite{tmQCD.Frezzotti.lat01,tmQCD.McNeile.lat01}.
%
It will be interesting to apply this action
to the non-perterburtive determination of improvement and 
renormalization constants at strong coupling,
where exceptional configurations cause
a serious problem\cite{NP-csw.ALPHA.Nf0,NP-Z.RG}.

\subsection{Negative parity baryon}

An interesting subject in recent quenched simulations 
is to explore the rich structure of the excited baryons.
There already exist several studies 
for the negative parity baryon $N^*$ ($J^{P}\!=\!1/2^-$)
at finite lattice spacings\cite{Nf0.N-}.
The first systematic study of scaling violation 
and finite size effects 
has been performed this year by the QCDSF-UKQCD-LHPC collaboration
using the $O(a)$-improved Wilson action\cite{QCDSF.Nf0.N-}.
Their result $m_{N^*}/m_{N}\!=\!1.50(3)(8)(15)$ 
in the continuum limit is consistent with the experimental value 1.63.

It is also interesting to explore other excitations.
A preliminary result
for the excited decuplet baryons, $J^{P}\!=\!3/2^+$ and $3/2^-$,
was presented by Lee {\it et al.} at this conference\cite{Lee.Nf0.N-}.

%
%

\section{Strong coupling constant}
\label{sec:alpha_s}


Recent results of $\alpha_{\MSbar}^{(N_f=5)}(M_z)$ 
from lattice QCD are summarized in Table~\ref{tab:alpha_s}.
%
%
%
Recent values using the Wilson-type quark action for sea 
quarks\cite{alphas.SESAM,alphas.UKQCD} are significantly smaller
compared to a standard lattice result previously obtained by Davies 
{\it et al.}\cite{alphas.NRQCD} using KS fermions.
%
%
In particular this year's result of the QCDSF-UKQCD collaboration
is smaller despite a consistent use of two-loop perturbation theory 
and a continuum extrapolation\cite{alphas.QCDSF}.
%
%

The origin of 
the difference between results with the KS and Wilson fermions 
should be clarified.  
Since the discrepancy is already present 
in the value of $\alpha_P$ at the cutoff scale in two-flavor QCD,
systematic uncertainties such as scaling violation and higher order 
corrections in perturbation theory 
should be studied carefully in future.

Calculations of $\alpha_s$ through various definitions
are also helpful to resolve this problem.
%
%
Boucaud {\it et al.} presented a preliminary result 
from lattice calculations of the gluon green functions\cite{alphas.Boucaud}.
A non-perturbative determination of $\alpha_s$ 
in the Schr\"odinger functional scheme\cite{alphaSF} 
was performed by ALPHA\cite{alphas.ALPHA} in two-flavor QCD,
and determination of the lattice energy scale 
is in progress\cite{alphas.ALPHA}.

\section{Quark masses}
\label{sec:mq}

By last year results for the light quark mass $m_{ud}$ 
in quenched QCD converged within an accuracy of about 
15\%\cite{review.mq}.  For $m_s$ there turned out an additional 
uncertainty of about 20\% 
due to the choice of the input meson mass
({\it e.g.} $m_K$ or $m_{\phi}$)\cite{CP-PACS.Nf0}.
A noticable trend in this year's calculation 
is the use of the domain-wall and overlap fermions.
Calculations of the charm quark mass $m_c$ 
have also been pursued in quenched QCD. 

In $N_f\!=\!2$ full QCD,
CP-PACS reported results of a systematic study already last year 
\cite{mq.CP-PACS}.
They found that sea quark effects reduce $m_{ud}$ and $m_s$ 
by about 25\%, 
and the discrepancy in $m_s$ with $m_K$ or $m_{\phi}$ as input
is remarkably reduced.
This year there has been less progress in full QCD.

\begin{table}[t]
\setlength{\tabcolsep}{0.2pc}
\caption{
  Recent results of $\alpha_{\MSbar}^{(N_f=5)}(M_z)$.
  All results except Ref.\cite{alphas.Boucaud} are calculated 
  through $\alpha_P$\cite{alphaP}. 
}
\begin{tabular}{lll}
\hline 
\hline 
group   &   quark   & $\alpha_{\MSbar}^{(N_f\!=\!5)}$  \\
\hline 
SESAM\cite{alphas.SESAM}       & Wilson     & 0.1118(17)       \\
QCDSF-UKQCD\cite{alphas.QCDSF} & clover(NP) & 0.1076(20)(18)   \\
Boucaud {\it et al.}\cite{alphas.Boucaud} & Wilson & 0.113(3)(4)\\
\hline 
Davies {\it et al.}\cite{alphas.NRQCD} & KS & 0.1174(24)  \\
\hline
\hline
\end{tabular}
\label{tab:alpha_s}
\vspace*{-5mm}
\end{table}

\subsection{Strange quark mass in quenched QCD}

A new estimate of $m_s$ was obtained by CP-PACS 
using the domain-wall fermion\cite{ms.CP-PACS}.
Their calculations were performed at $a^{-1}\!\sim\!2$ and 3~GeV
using one-loop perturbative renormalization constants.
%
%
Giusti {\it et al.} calculated $(m_{s}+m_{ud})$ 
at $a^{-1}\!\sim\!2$~GeV using the overlap fermion\cite{ms.Giusti}.
Their renomalization constants are evaluated 
non-perturbatively in the RI/MOM scheme.
%
%

\begin{figure}[t]
\vspace*{-1mm}
\begin{center}
   \leavevmode
   \includegraphics*[width=72mm,clip]{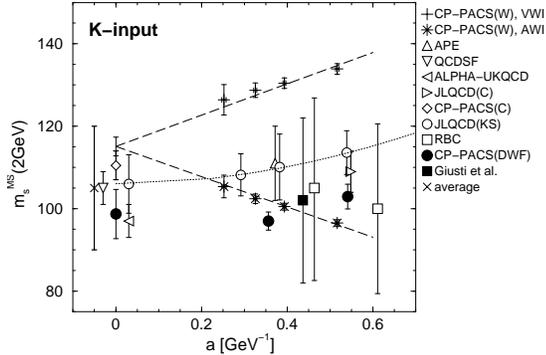}
\end{center}
\vspace*{-15mm}
\caption{
  Continuum extrapolation of strange quark mass
  with $m_K$ as input in $\MSbar$ scheme at scale $\mu\!=\!2$~GeV.
  Filled and open symbols show new and previous results,
  respectively.
  The choice of the quark action is written in brackets,
  where ``W''=Wilson, ``C''=clover, ``KS''=Kogut-Susskind and 
  ``DWF''=domain-wall fermions.
}
\vspace*{-6mm}
\label{fig:mq:ms}
\end{figure}

Recent results with $m_K$ as input are plotted 
in Fig.~\ref{fig:mq:ms}.
A reasonable agreement among the new and previous results 
is observed.
Therefore, the average 
$m_{s}^{\MSbar}(\mbox{2~GeV})=105 \pm 15~{\rm MeV}$
does not change significantly since last year\cite{review.mq}.


\subsection{Light quark mass in quenched QCD}

We consider the ratio $m_s/m_{ud}$, 
rather than $m_{ud}$ itself, since 
systematic uncertainties partially cancel in this ratio.
A new result $m_s/m_{ud}=26.2(2.3)$ by CP-PACS\cite{ms.CP-PACS} 
as well as previous results are in good agreement with 
the prediction of one-loop ChPT :  24.4(1.5)\cite{mq.ChPT}.
We therefore apply the ratio from ChPT to the average of $m_s$ 
to obtain 
$m_{ud}^{\MSbar}(\mbox{2~GeV})=4.3 \pm 0.7~{\rm MeV}$.

Dong {\it et al.} calculated $m_{ud}$ 
with the overlap fermion\cite{mud.Dong}.
Their estimate $m_{ud}^{\MSbar}(\mbox{2~GeV})\!=\!5.3(0.6)(0.7)$~MeV
agrees with the above average.

\subsection{Charm quark mass in quenched QCD}

There have been few studies for calculating 
the charm quark mass $m_c$\cite{mc.previous}.
The reason is that $O((am_q)^n)$ scaling violation
is expected to be large in relativistic formulations,
while the charm quark is too light 
to be treated in the non-relativistic approximation.

New estimates of $m_c$ have been reported by two groups this year, both 
using the $O(a)$-improved action to reduce scaling violation.
Becirevic {\it et al.} performed a simulation 
at a fixed lattice spacing $a^{-1}\!\sim\!3$~GeV\cite{mc.Becirevic}.
The results significantly differ for $m_c$ obtained from vector(VWI) 
and axial vector Ward identities(AWI).
They take an average 
to quote $m_c^{\MSbar}(m_c)\!=\!1.26(3)(12)$. 
The difference between the two methods(AWI and VWI) and the average 
is treated as a systematic error. 
%
%
%

ALPHA performed simulations
in a wide range of lattice spacing $a^{-1}\!\sim\!2$--4~GeV 
and performed the continuum extrapolation\cite{mc.ALPHA}.
They observed a nice convergence of $m_c$ 
determined from AWI and VWI toward the continuum limit,
and obtained $m_c^{\MSbar}(m_c)\!=\!1.314(40)(20)(7)$ 
in the continuum limit.

\begin{figure}[t]
\vspace*{-4.5mm}
\begin{center}
   \leavevmode
   \includegraphics*[width=59mm,clip]{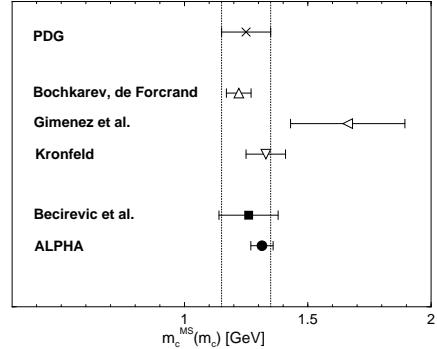}
\end{center}
\vspace*{-16mm}
\caption{
  Recent results of charm quark mass.
  An average quoted by PDG\cite{PDG}
  is also plotted.
}
\vspace*{-6mm}
\label{fig:mq:mc}
\end{figure}

These results are plotted in Fig.~\ref{fig:mq:mc}  
together with previous results\cite{mc.previous} 
and an average by PDG\cite{PDG}.
The two new results are consistent with previous ones
as well as the PDG's average.

A new estimate using the heavy quark action in the FNAL interpretation 
\cite{Fermilab} 
was presented by Juge {\it et al.} at this conference\cite{mc.Juge}.
While the estimation of the systematic error is still in progress, 
their preliminary result is also around 1.3~GeV 
and consistent with the above new results.

\subsection{Results in $N_f\!=\!2$ full QCD}

SESAM-T$\chi$L 
attempted a continuum extrapolation 
of $m_{ud}$ and $m_s$ from previous\cite{mq.SESAM.prev}
and new simulations at $\beta\!=\!5.5$\cite{SESAM.Nf2.beta55}. 
Their range of the lattice spacing is, however, narrow
and far from the continuum limit.
This leads to a large uncertainty in their results,
$m_{ud}^{\MSbar}(2~{\rm GeV})\!=\!4.5(1.7)$~MeV and 
$m_{s}^{\MSbar}(2~{\rm GeV})\!=\!92(83)$~MeV.
In near future, 
we expect results from the UKQCD and JLQCD simulations
with the $O(a)$-improved Wilson fermion.

\section{Conclusions}
\label{sec:conclusion}


In quenched QCD, improved formulations,  
such as the domain-wall and overlap fermions 
and anisotropic lattice actions,
have been tested by calculating 
the light hadron spectrum and quark masses,
and good consistency with previous results has been observed.
These formulations are ready for applications
to precise determinations of other observables.

In $N_f\!=\!2$ full QCD,
sea quark effects on the meson spectrum,
which were previously observed by CP-PACS,
have been confirmed using 
the $O(a)$-improved Wilson and an improved KS fermions.
However we still need further studies 
for sea quark effects on the baryon and glueball spectrum.
Exploring small sea quark masses with Wilson-type fermions
is also an important issue 
to resolve the poor consistency between lattice data and ChPT.

Pursuit of realistic simulations in three flavor QCD 
is one of the most important issue in spectroscopic studies.
This year significant progress to this end has been made 
by several groups using both of the KS and Wilson fermions.
Further systematic studies will be expected in near future.
   
And finally, there has been few progress in non-perturbative 
calculation of the improvement coefficient $c_A$ and renormalization 
factors in full QCD.
These calculations are urgently required for precise 
determinations of the quark masses and decay constants.

\vspace{-2mm}
\section*{Acknowledgments}

I thank   
   Y.~Aoki,  G.S.~Bali, C.~Bernard, S.~Collins, 
   G.~Di~Carlo, S.J.~Dong, N.~Eicker, 
   R.~Frezzotti, C.~Gebert, L.~Giusti,
   A.~Hart, S.~Hashimoto,  S.~Hauswirth, J.~Hein, 
   K.~Ide, A.C.~Irving, K-I.~Ishikawa, 
   K.~Jansen, 
   F.X.~Lee, L.~Levkova, X.Q.~Luo, T.~Manke, 
   H.~Matsufuru, C.~McNeile, H.~Moutarde, A.~Nakamura, 
   M.~Okawa, K.~Orginos, O.~Philipsen, 
   C.~Rebbi, J.~Rolf, G.~Rossi, S.~Sint, 
   H.~Wittig, T.~Yamazaki, and A.~Ukawa
for communications and useful discussions.
I also thank 
  S.~Hashimoto, Y.~Iwasaki, M.~Okawa and A.~Ukawa
for valuable suggestions on the manuscript.
This work is supported by the JSPS Research Fellowship.


\end{document}